\documentclass[12pt,a4paper]{article}

\newcommand{\e}{\epsilon}
\newcommand{\ta}{\theta} 
\newcommand{\de}{\Delta x^2}

\begin{document}

\begin{flushright}
hep-th/9909186
\end{flushright}
\vspace{1.8cm}

\begin{center}
 \textbf{\Large The Hadamard Function and the Feynman Propagator 
 \\ in the AdS/CFT Correspondence}
\end{center}
\vspace{1.6cm}
\begin{center}
 Shijong Ryang
\end{center}

\begin{center}
\textit{Department of Physics \\ Kyoto Prefectural University of Medicine
\\ Taishogun, Kyoto 603-8334 Japan}
\par
\texttt{ryang@koto.kpu-m.ac.jp}
\end{center}
\vspace{2.8cm}
\begin{abstract}
We construct the retarded Green function and the Hadamard function in the
Lorentzian (d+1)-dimensional anti-de Sitter spacetime for the Poincar\'e
coordinate by performing the mode integration directly. We explore the 
structure of singularities for the position-space Feynman propagator 
derived from them. The boundary scaling limits of the bulk Feynman
propagator yield the bulk-boundary propagator and the boundary conformal
correlation function with an extra factor.
\end{abstract}
\vspace{3cm}
\begin{flushleft}
 September, 1999
\end{flushleft}
\newpage
The Maldacena conjecture has brought a renewed interest in the 
relationship between the large $N$ Yang-Mills theory and the string 
theory \cite{JM}. In particular the semiclassical limit of IIB 
superstring theory on $AdS_5\times S_5$ is related to the large $N$ limit
of $\mathcal{N}=4$ super Yang-Mills theory. This AdS/CFT 
correspondence \cite{JM,GKP,EW} between the quantum gravity in the 
anti-de Sitter space and the conformal gauge theory on the boundary has 
been considered in the Euclidean space. The correlation functions for the
boundary conformal theory have been computed from the bulk dynamics in
the AdS space \cite{MV,FMMR,LH}.

The AdS/CFT correspondence in the Lorentzian 
spacetime and scattering processes have been investigated 
\cite{BDHM,BKLT,EKV,CS}.
There are trials to extract the local and causal gravity dynamics in the
Lorentzian AdS spacetime from the boundary conformal field theory 
\cite{HI,JP,LS,BGL,SG,BPLB}.
From this view point the conformal correlation function is interpreted as
the S-matrix in the flat space limit \cite{JP,LS} or some kind of the 
S-matrix \cite{BGL,SG}. The bulk massive scalar field in the Lorentzian 
AdS spacetime is composed of the normalizable propagating modes and the 
non-normalizable modes whose boundary values act as sources for conformal
operators in the boundary theory \cite{BKLT}. The normalizable
modes are quantized in the curved AdS spacetime and vanish at the 
boundary. They correspond to states in the Hilbert space of the boundary 
conformal theory. Generally there are many Green functions in a Lorentzian
curved spacetime. For the AdS spacetime the Feynman propagator 
of the massive scalar field in the momentum
space has been presented as a sum of these normalizable modes \cite{BGL}.
With respect to the retarded Green function the configuration form in the
$(d+1)$-dimensional AdS spacetime for the Poincar\'e coordinate
has been derived by carrying out the mode integration in the lower
dimensions and taking advantage of a recursion relation between the
$AdS_{d-1}$ Green function and the $AdS_{d+1}$ one \cite{DVK}.

For the Lorentzian two-dimensional AdS spacetime the Hadamard functions 
in the configuration space have been constructed by summing the 
normalizable modes about various vacua which correspond to  
selections of the  time coordinate \cite{SS}. 
The Hadamard function in the global vacuum or the
Poincar\'e vacuum is expressed in terms of the SL(2,R)-invariant distance
variable, while in the Boulware vacuum it is not so expressed.
The short-distance behaviors of the Hadamard functions of massive 
scalars  are investigated and the massless limits of
them are shown to yield the usual massless Green functions on the strip
or the half plane that are described by logarithmic functions.
From the short-distance behaviors the vacuum expectation values 
of the stress tensors are studied in the various $AdS_{1+1}$ vacua.

We will try to construct the Hadamard function in the Lorentzian 
$(d+1)$-dimensional AdS spacetime for the
Poincar\'e coordinate. For that purpose, first in the momentum
representation of the retarded Green function  
we will  perform the mode integration directly
in $AdS_{d+1}$ by using a recursion relation of the Bessel functions.
We will use this direct method to determine the position-space 
Hadamard function, which will be compactly described in the differential
expressions. Combining them we will 
derive the Feynman propagator, which is 
represented in the configuration space so that we can analyse what kinds
of singularities appear. We will demonstrate what happens in the 
$(d+1)$-dimensional bulk Feynman propagator containing the Heaviside 
functions, when one argument is taken to the boundary. In this boundary
scaling limit the bulk-boundary propagator of the Lorentzian signature
version will be extracted. 

We consider a massive scalar field in the Lorentzian $(d+1)$-dimensional
AdS spacetime expressed by the Poincar\'e coordinate 
$(t,\vec{x},z) = (x,z), ds^2 = R^2/z^2 (-dt^2 +d\vec{x}^2 + dz^2 ),$
where $(t,\vec{x})$ spans the $d$-dimensional Minkowski spacetime and 
$0 \le z \le \infty$ with a boundary at $z = 0$ and a horizon at 
$z = \infty$. The general regular solution of the bulk wave equation is
sum of a fluctuating normalizable mode solution, whose existence is a
feature of Lorentzian signature, and a non-normalizable mode solution
\cite{BKLT}. Near the boundary the propagating mode behaves as $z^{2h_+}$,
while the non-fluctuating mode as $z^{2h_-}$ with $2h_{\pm} = d/2 \pm
\nu = d/2 \pm \sqrt{m^2R^2 + d^2/4}$ where $m$ is mass of the scalar 
field. Hereafter we will set $R = 1$. For the Poincar\'e vacuum the 
propagating bulk mode solution is expanded as
\begin{equation}
\Phi(z,x) =\int dqd^{d-1}\vec{k} z^{\frac{d}{2}}J_{\nu}(qz)
\left(\frac{q}{2(2\pi)^{d-1}\omega}\right)^{\frac{1}{2}} 
(e^{-i\omega t + i\vec{k}\cdot \vec{x}}\hat{b}_{q\vec{k}} + \mathrm{h}
.\mathrm{c}.)
\label{mod}\end{equation}
with $\omega^2 = q^2 + k^2$ \cite{BGL}. 
From the small $z$ expansion of the
Bessel functions $J_{\nu}(qz)$ we see that the normalizable mode indeed
vanishes as $\Phi \sim z^{d/2 +\nu}$ 
at the boundary. The annihilation operators 
$\hat{b}_{q\vec{k}}$ define the Poincar\'e vacuum $|0>$ as 
$\hat{b}_{q\vec{k}}|0>=0$ and obey the commutation 
relations with the creation 
operators $\hat{b}^{\dagger}_{q\vec{k}}, \; [ \hat{b}_{q\vec{k}},
\hat{b}^{\dagger}_{q'\vec{k}'}] = \delta(q-q')\delta^{d-1}
(\vec{k}-\vec{k}')$. 

We begin with the retarded Green function defined by
$G_R(z,x;z',x') = - \e(t-t')/2i \,<0|[\Phi(z,x),\Phi(z',x')]|0>.$
Through the commutation relations it can be expressed as
\begin{equation}
-\frac{1}{2}\e(t-t')\int_0^{\infty}dqq(zz')^{d/2}J_{\nu}(qz)J_{\nu}(qz')
\Delta(x,x')
\label{grt}\end{equation}
with $i\Delta(x,x')=\int dk_0 d^{d-1}\vec{k}/(2\pi)^{d-1}\e(k_0)\delta
(k_0^2-\omega^2)\exp [-ik_0(t-t')+ i\vec{k}\cdot (\vec{x}-\vec{x'})].$
For the $AdS_{2+1}$ case by carrying out an integration \cite{GR}
\begin{eqnarray}
\int_0^{\infty}dq q J_{\nu}(qz)J_{\nu}(qz')J_0(qs) =
\sqrt{\frac{2}{\pi^3}}\frac{1}{zz'}[-i\sin\nu\pi\ta(s-z -z')
(U^2-1)^{-\frac{1}{4}}Q^{\frac{1}{2}}_{\nu-\frac{1}{2}}(U) \nonumber \\
+ \frac{\pi}{2}\ta(z+z'-s)\ta(s-|z-z'|)
(1-V^2)^{-\frac{1}{4}}P^{\frac{1}{2}}_{\nu-\frac{1}{2}}(V)],
\label{jin}\end{eqnarray}
where $s=\sqrt{(t-t')^2 - (x-x')^2}=\sqrt{-\de}, U = (-\de -z^2-z'^2)
/2zz'$ and $V = (\de + z^2 + z'^2)/2zz',$ we have
\begin{equation}
G_R = \frac{\ta(-\de)}{(2\pi)^{\frac{3}{2}}}[-i\sin \nu\pi\ta(U-1)
(U^2-1)^{-\frac{1}{4}}Q^{\frac{1}{2}}_{\nu-\frac{1}{2}}(U)
+\frac{\pi}{2}\ta(1-|V|) (1-V^2)^{-\frac{1}{4}}
P^{\frac{1}{2}}_{\nu-\frac{1}{2}}(V)].
\label{rtw}\end{equation}
Here $P^{\frac{1}{2}}_{\nu-\frac{1}{2}},Q^{\frac{1}{2}}_{\nu-\frac{1}{2}}$
are the associated Legendre functions of the first and second kinds.
Next the $AdS_{3+1}$ case requires an integration of 
$\int dqq^{3/2}J_{-1/2}(qs)J_{\nu}(qz)J_{\nu}(qz')$, however which is
divergent. Through a recursion relation $J_{-1/2}(u) = (d/du
+ 1/2u)J_{1/2}(u)$ it becomes $(d/ds+1/2s)\int dq
q^{1/2}J_{1/2}(qs)J_{\nu}(qz)J_{\nu}(qz').$ This finite integration 
\cite{GR} leads to
\begin{eqnarray}
G_R = \frac{\ta(-\de)}{4\pi^2}\frac{zz'}{s}[\sin(\nu-\frac{1}{2})\pi
(\delta(s-z-z')Q_{\nu-\frac{1}{2}}(U) +\ta(s-z-z')
\frac{dQ_{\nu-\frac{1}{2}}(U)}{ds})  \nonumber \\ + 
\frac{\pi}{2}((\delta(s-|z-z'|) - 
\delta(z+z'-s))P_{\nu-\frac{1}{2}}(V) + \ta(z+z'-s)\ta(s-|z-z'|)
\frac{dP_{\nu-\frac{1}{2}}(V)}{ds})],
\label{rth}\end{eqnarray}
where $P_{\nu-\frac{1}{2}}, Q_{\nu-\frac{1}{2}}$ are the Legendre 
functions of the first and second kinds. This differential expression with
respect to $s$ as well as (\ref{rtw}) are identical with those presented 
in Ref. \cite{DVK}. It can be further 
simplified in terms of $U$ and $V$ as
\begin{equation}
G_R = \frac{\ta(-\de)}{4\pi^2}[\sin(\nu-\frac{1}{2})\pi\frac{d}{dU}
(\ta(U-1)Q_{\nu-\frac{1}{2}}(U)) - \frac{\pi}{2}\frac{d}{dV}
(\ta(1-|V|)P_{\nu-\frac{1}{2}}(V))].
\label{rthu}\end{equation}
In deriving this compact expression we have used $\ta(1-|V|)=\ta(1-V)
\ta(1+V)$ and $\frac{d}{dV}\ta(1-\sqrt{V^2}) = -\e(V)\delta(1-\sqrt{V^2})
=\delta(1+V) - \delta(1-V)$ and observed that the factor $zz'$ is 
canceled. We have used  the recursion relation of Bessel functions, which
prescription will be extended in considering the general $AdS_{d+1}$ case.
In view of $U = -1-\sigma, V = 1+\sigma,$ the retarded Green function is 
described as a function of the AdS-invariant distance variable 
$\sigma = (\de + (z-z')^2)/2zz'$. Moreover we proceed to determine $G_R$
in $AdS_{4+1}$. Substituting the usual four-dimensional invariant
$\Delta$ function $\Delta = -\e(t-t')/2\pi[\delta(\de) - q\ta(-\de)
J_1(q\sqrt{-\de})/2\sqrt{-\de}]$ into (\ref{grt})
we note that the first term is evaluated by using
\begin{equation}
\int_0^{\infty}dqqJ_{\nu}(qz)J_{\nu}(qz')
=\frac{\delta(z-z')}{\sqrt{zz'}}
\label{jde}\end{equation}
however, the second term is so divergent that $J_1(q\sqrt{-\de})$ should
be replaced by $\frac{\sqrt{-\de}}{qr}\frac{\partial J_0}{\partial r}$
with $r =|\vec{x}-\vec{x'}|$. Applying of (\ref{jin}) yields an involved
expression
\begin{eqnarray}
G_R = -\frac{1}{8\pi} [ -2(zz')^\frac{3}{2}\delta(\de)\delta(z-z') +
\ta(-\de)\frac{i\sin\nu\pi}{\sqrt{\pi^3/2}}\frac{d}{dU}(\ta(U-1)
\frac{Q^{\frac{1}{2}}_{\nu-\frac{1}{2}}(U)}{(U^2-1)^{1/4}}) \nonumber \\
+ \frac{\ta(-\de)}{\sqrt{2\pi}}((\delta(V+1)-\delta(V-1))
\frac{P^{\frac{1}{2}}_{\nu-\frac{1}{2}}(V)}{(1-V^2)^{1/4}} +\ta(1-|V|)
\frac{d}{dV}\frac{P^{\frac{1}{2}}_{\nu-\frac{1}{2}}(V)}{(1-V^2)^{1/4}})].
\label{rfo}\end{eqnarray}

In the following, we attempt a general derivation of the retarded Green 
function in $AdS_{d+1}$. Performing the integrations over $k_0$ and 
angular variables in $\Delta (x,x')$ 
of (\ref{grt}) leaves the radial integration over
$k=|\vec{k}|, \; \int_0^{\infty}dkk^{\frac{d-1}{2}}J_{\frac{d-3}{2}}(kr)
\sin(\sqrt{k^2+q^2}(t-t'))/\sqrt{k^2+q^2}$ which is not convergent.
Using the recursion relations of Bessel functions we construct interesting
relations for $d =$ odd
\begin{equation}
J_{\frac{d-3}{2}}(kr)=\frac{1}{k^{\frac{d-3}{2}}}D(\frac{d-5}{2})
D(\frac{d-7}{2})\cdots D(1)D(0)J_0(kr)
\label{ito}\end{equation}
with $D(a) = -d/dr + a/r$, and for $d =$ even
\begin{equation}
J_{\frac{d-3}{2}}(kr)=\frac{1}{k^{\frac{d-2}{2}}}D(\frac{d-5}{2})
D(\frac{d-7}{2})\cdots D(\frac{1}{2})D(-\frac{1}{2})
J_{-\frac{1}{2}}(kr).
\label{ite}\end{equation}
These relations make the radial integration finite. For even $d$
the finite radial integral becomes $\e(t-t')\sqrt{\frac{\pi}{2}}
D(\frac{d-5}{2})\cdots D(-\frac{1}{2})(\ta(-\de)r^{-1/2}
J_0(q\sqrt{-\de})).$ We again use (\ref{jin}) to obtain a formal
expression
\begin{eqnarray}
G_R=\frac{1}{(2\pi)^{\frac{d+1}{2}}} \frac{(zz')^{\frac{d}{2}-1}}
{r^{\frac{d-3}{2}}}D(\frac{d-5}{2})\cdots D(-\frac{1}{2})\ta(-\de)
r^{-\frac{1}{2}}[-i\sin\nu\pi \nonumber \\
\times \ta(U-1)(U^2-1)^{-\frac{1}{4}}Q^{\frac{1}{2}}_{\nu-\frac{1}{2}}(U)
+ \frac{\pi}{2}\ta(1-|V|)(1-V^2)^{-\frac{1}{4}}
P^{\frac{1}{2}}_{\nu-\frac{1}{2}}(V)].
\end{eqnarray}
Owing to a factor $r^{-1/2}$, for instance $D(\frac{1}{2})
D(-\frac{1}{2})r^{-1/2}f(U)$ is changed into a simple form $r^{-1/2}
(r/zz')^2\frac{d^2f}{dU^2}$. Therefore the complicated formal expression
is to some extent simplified as
\begin{eqnarray}
G_R &=& \frac{1}{(2\pi)^{\frac{d+1}{2}}}[-i\sin\nu\pi (\frac{d}{dU})^{
\frac{d-2}{2}}(\ta(-\de)\ta(U-1)(U^2-1)^{-\frac{1}{4}}
Q^{\frac{1}{2}}_{\nu-\frac{1}{2}}(U)) \nonumber \\
&+& \frac{\pi}{2}(-\frac{d}{dV})^{\frac{d-2}{2}}(\ta(-\de)\ta(1-|V|)
(1-V^2)^{-\frac{1}{4}}P^{\frac{1}{2}}_{\nu-\frac{1}{2}}(V))],
\label{rde}\end{eqnarray}
which is real because of imaginary $Q^{\frac{1}{2}}_{\nu-\frac{1}{2}}(U)$
and real $P^{\frac{1}{2}}_{\nu-\frac{1}{2}}(V)$. Turning to the odd $d$ 
case we have the radial integral  $\e(t-t')\sqrt{\frac{\pi}{2}}
D(\frac{d-5}{2})\cdots D(0)(\ta(-\de)(q/\sqrt{-\de})^{1/2}J_{-1/2}
(q\sqrt{-\de}))$ where $J_{-1/2}(qs)$ is replaced by $q^{-1}
(d/ds + 1/2s)J_{1/2}(qs)$. We again obtain a formal
expression
\begin{eqnarray}
G_R=\frac{1}{(2\pi)^{\frac{d+1}{2}}} (\frac{zz'}{r})^{\frac{d-3}{2}}
D(\frac{d-5}{2})\cdots D(0)\ta(-\de)
[\sin(\nu-\frac{1}{2})\pi \nonumber \\
\times \frac{d}{dU}(\ta(U-1)Q_{\nu-\frac{1}{2}}(U))
- \frac{\pi}{2}\frac{d}{dV}(\ta(1-|V|)P_{\nu-\frac{1}{2}}(V))],
\end{eqnarray}
which is also simplified as 
\begin{eqnarray}
G_R &=& \frac{1}{(2\pi)^{\frac{d+1}{2}}} [ \sin(\nu-\frac{1}{2})\pi
(\frac{d}{dU})^{\frac{d-3}{2}}(\ta(-\de)\frac{d}{dU}
(\ta(U-1)Q_{\nu-\frac{1}{2}}(U))) \nonumber \\
&-& \frac{\pi}{2}(-\frac{d}{dV})^{\frac{d-3}{2}}(\ta(-\de)\frac{d}{dV}
(\ta(1-|V|)P_{\nu-\frac{1}{2}}(V)))],
\label{rdo}\end{eqnarray}
whose two terms are real. This result, when $d = 3$, becomes (\ref{rthu}).
For $d=2$ the expression (\ref{rde}) leads to (\ref{rtw}), while for 
$d=4$ the agreement between (\ref{rde}) and (\ref{rfo}) can be shown 
through a relation which is obtained 
by putting $\de = 0$ in (\ref{jin}) and combining with (\ref{jde}).
For massless scalars $\nu=d/2$ the retarded Green functions, (\ref{rde})
and (\ref{rdo}) are further simplified to be described only by 
$P^{\frac{1}{2}}_{\nu-\frac{1}{2}}(V)$ and $P_{\nu-\frac{1}{2}}(V)$
respectively. 

Here based on above the prescription we are able to construct the Hadamard
function $G^{(1)}(z,x;z',x') = <0|\{ \Phi(z,x), \Phi(z',x')\}|0>$ in 
$AdS_{d+1}$. Substituting the mode expansion (\ref{mod}) we have
\begin{equation}
G^{(1)}=\int_0^{\infty}dqq(zz')^{\frac{d}{2}}J_{\nu}(qz)J_{\nu}(qz')
\int\frac{d^{d-1}\vec{k}}{(2\pi)^{d-1}}
\frac{e^{i\vec{k}\cdot(\vec{x}-\vec{x'})}}{\sqrt{k^2+q^2}}
\cos \sqrt{k^2+q^2}(t-t').
\end{equation}
In the integration over $\vec{k}$ performing the angular  
integrations gives rise to a radial one 
$\int_0^{\infty}dkk^{\frac{d-1}{2}}
J_{\frac{d-3}{2}}(kr)\cos(\sqrt{k^2+q^2}(t-t'))/\sqrt{k^2+q^2}$
which is divergent. For even $d$, $J_{\frac{d-3}{2}}(kr)$ is replaced
by $J_{-\frac{1}{2}}(kr)$ through the successive relation (\ref{ite}) so 
that the radial integration can be carried out as 
$D(\frac{d-5}{2})\cdots D(-\frac{1}{2})r^{-1/2}[-\ta(-\de)
\sqrt{\frac{\pi}{2}}N_0(q\sqrt{-\de}) + \ta(\de)\sqrt{\frac{2}{\pi}}
K_0(q\sqrt{\de})]$ where $N_0$ is the Neumann function and $K_0$ is the
modified Bessel function of the second kind. Using a formula in 
Ref. \cite{PBM} for $\int dqqJ_{\nu}(qz)J_{\nu}(qz')N_0(q\sqrt{-\de})$
we arrive at 
\begin{eqnarray}
G^{(1)} &=&-\frac{2}{(2\pi)^{\frac{d+1}{2}}}[-i\cos\nu\pi 
(\frac{d}{dU})^{\frac{d-2}{2}}\ta(-\de)\ta(U-1)(U^2-1)^{-\frac{1}{4}}
Q^{\frac{1}{2}}_{\nu-\frac{1}{2}} \nonumber \\
&+& (-\frac{d}{dV})^{\frac{d-2}{2}}\ta(-\de)(-\ta(1-|V|)
(1-V^2)^{-\frac{1}{4}}Q^{\frac{1}{2}}_{\nu-\frac{1}{2}}+
i\ta(V-1)(V^2-1)^{-\frac{1}{4}}Q^{\frac{1}{2}}_{\nu-\frac{1}{2}})
 \nonumber \\
&+&i(-\frac{d}{dV})^{\frac{d-2}{2}}\ta(\de)(V^2-1)^{-\frac{1}{4}}
Q^{\frac{1}{2}}_{\nu-\frac{1}{2}}],
\label{hae}\end{eqnarray}
where each argument of $Q^{\frac{1}{2}}_{\nu-\frac{1}{2}}$ has been 
suppressed, however every term is real. The $d =$ odd case requires
that $J_{\frac{d-3}{2}}(kr)$ is put backward to  $J_{0}(kr)$ through
(\ref{ito}). The radial integration can be carried out to be
$D(\frac{d-5}{2})\cdots D(0)q^{1/2}[-\ta(-\de)(\sqrt{-\de})^{-1/2}
\sqrt{\frac{\pi}{2}}N_{-1/2}(q\sqrt{-\de}) + \ta(\de)(\sqrt{\de})^{-1/2}
\sqrt{\frac{2}{\pi}}K_{1/2}(q\sqrt{\de})]$. But 
the first term of the remaining integration given by
$\int dqq^{3/2}J_{\nu}(qz)J_{\nu}(qz')N_{-1/2}(q\sqrt{-\de})$ is so 
divergent that $N_{-1/2}(qs)$ should be replaced by $q^{-1}(d/ds
+1/2s)N_{1/2}(qs)$. The integration $\int dqq^{1/2}J_{\nu}(qz)
J_{\nu}(qz')N_{1/2}(q\sqrt{-\de})$ \cite{PBM} produces a factor 
$(\sqrt{-\de})^{-1/2}$, which converts the operation $d/ds
+1/2s$ into $d/dU$ or $d/dV$, in the same way as
the step to (\ref{rth}), (\ref{rthu}). Making the successive operations
of $D(a)$ we derive the Hadamard function 
\begin{eqnarray}
G^{(1)} &=& -\frac{2}{(2\pi)^{\frac{d+1}{2}}}[\cos(\nu-\frac{1}{2})\pi 
(\frac{d}{dU})^{\frac{d-3}{2}}(\ta(-\de)\frac{d}{dU}
(\ta(U-1)Q_{\nu-\frac{1}{2}})) \nonumber \\
&+& (-\frac{d}{dV})^{\frac{d-3}{2}}(\ta(-\de)\frac{d}{dV}
(\ta(1-|V|)Q_{\nu-\frac{1}{2}}+\ta(V-1)Q_{\nu-\frac{1}{2}})) \nonumber \\
&+& (-\frac{d}{dV})^{\frac{d-3}{2}}(\ta(\de)\frac{d}{dV}
Q_{\nu-\frac{1}{2}})],
\label{hao}\end{eqnarray}
whose every term is real. Thus we construct the Hadamard function in
$AdS_{d+1}$ expressed in terms of the invariant distance variable 
$\sigma$. The last two terms in (\ref{hae}) and (\ref{hao}) show 
contributions from the spacelike separation. The timelike region is
composed of two parts which are characterized by $\ta(U-1)$ and 
$\ta(1-|V|)$. 

Gathering together we can obtain the Feynman propagator $iG_F = 
G_R + \frac{i}{2}G^{(1)}$ in $AdS_{d+1}$. For even $d$
it is given by
\begin{eqnarray}
iG_F &=& \frac{1}{(2\pi)^{\frac{d+1}{2}}}[-e^{i\nu\pi} 
(\frac{d}{dU})^{\frac{d-2}{2}}\ta(-\de)\ta(U-1)(U^2-1)^{-\frac{1}{4}}
Q^{\frac{1}{2}}_{\nu-\frac{1}{2}} \nonumber \\
&+& (-\frac{d}{dV})^{\frac{d-2}{2}}\ta(-\de)\ta(1-|V|)
(1-V^2)^{-\frac{1}{4}}(\frac{\pi}{2}P^{\frac{1}{2}}_{\nu-\frac{1}{2}}+
iQ^{\frac{1}{2}}_{\nu-\frac{1}{2}}) \nonumber \\
&+& (-\frac{d}{dV})^{\frac{d-2}{2}}\ta(V-1)(V^2-1)^{-\frac{1}{4}}
Q^{\frac{1}{2}}_{\nu-\frac{1}{2}}], 
\label{sum}\end{eqnarray}
where an identity $\ta(\de)=\ta(\de)\ta(V-1)$ has been used.
It is noted that each first term in $G_R$ (\ref{rde}) and $G^{(1)}$
 (\ref{hae}) combines to yield a phase factor $e^{i\nu\pi}$ and
the combination $\frac{\pi}{2}P^{\frac{1}{2}}_{\nu-\frac{1}{2}}(V)+
iQ^{\frac{1}{2}}_{\nu-\frac{1}{2}}(V)$ is suggestively expressed as
$e^{-i\pi /4}Q^{\frac{1}{2}}_{\nu-\frac{1}{2}}(V+i\e)$.
This phase factor $e^{i\nu\pi}$ appears similarly in the Feynman 
propagator for odd $d$ 
\begin{eqnarray}
iG_F &=& \frac{1}{(2\pi)^{\frac{d+1}{2}}}[-e^{i\nu\pi} 
(\frac{d}{dU})^{\frac{d-3}{2}}\ta(-\de)\frac{d}{dU}
(\ta(U-1)Q_{\nu-\frac{1}{2}}) \nonumber \\
&-& (-\frac{d}{dV})^{\frac{d-3}{2}}\ta(-\de)\frac{d}{dV}
(\ta(1-|V|)(\frac{\pi}{2}P_{\nu-\frac{1}{2}} +
iQ_{\nu-\frac{1}{2}})) \nonumber \\
&-& i(-\frac{d}{dV})^{\frac{d-3}{2}}\ta(V-1)\frac{d}{dV}
Q_{\nu-\frac{1}{2}}],
\label{gfo}\end{eqnarray}
where the combination $\frac{\pi}{2}P_{\nu-\frac{1}{2}}(V) +
iQ_{\nu-\frac{1}{2}}(V)$ is also expressed as 
$iQ_{\nu-\frac{1}{2}}(V+i\e)$.

Now we are ready to compare the retarded Green function $G_R$ and the 
Feynman propagator $G_F$ in the Lorentzian $AdS_{d+1}$ spacetime. We note
that as shown in (\ref{rde}), (\ref{rdo}) $G_R$ is zero outside the 
light-cone. In $AdS_{2+1}, G_R$ (\ref{rtw}) is supported in the
timelike region $\sigma <0$ with no delta-function singularity, while 
$G_R$ (\ref{rthu}) in $AdS_{3+1}$ and $G_R$ (\ref{rfo}) in $AdS_{4+1}$
have contributions from not only the light-cone which contains coincident
points and is specified by $\delta(V-1)$, but also the timelike region
which is specified by $\ta(U-1)$ and $\ta(1-|V|)$. Moreover the latter
ones are singular on the reflected light-cone as expressed by 
$\delta(U-1)$ or $\delta(V+1)$, that is $\delta(\sigma + 2)$.
This second possible singularity occurs at $\de + (z+z')^2 = 0$,
which comes from an image field source at $z'_{I}=-z'$. On the other
hand as seen in the last term in (\ref{sum}) or (\ref{gfo})
$G_F$ has additional nonzero contribution from the spacelike region
specified by $V > 1$, which originates in $G^{(1)}$. 
The Feynman propagator in (\ref{sum}) for even $d$ is singular at  
$U =1$ and $V=\pm1$ in view of 
$Q^{\frac{1}{2}}_{\nu-\frac{1}{2}}(U)=i\sqrt{\pi/2}
(U^2-1)^{-\frac{1}{4}}(U+\sqrt{U^2-1})^{-\nu}$
for $U>1$, $P^{\frac{1}{2}}_{\nu-\frac{1}{2}}(V)=\sqrt{2/\pi}
(1-V^2)^{-\frac{1}{4}}\cos(\nu\cos^{-1}V)$ for $|V|<1$ and so on.
For odd $d$ case $Q_{\nu-\frac{1}{2}}$ in (\ref{gfo}) has logarithmical
singularities so that as $d$ becomes larger through increasing 
differentiations $G_F$ contains more various combinations of singularities
not only at the light-cone $V=1$, but also at the reflected light-cone
, $U=1$ and $V=-1$. It is interesting to note that the Feynman propagator
for the massless scalar $\nu=3/2$ in $AdS_{3+1}$ can be 
expressed in terms of a logarithmic function as 
\begin{eqnarray}
G_F &=& \frac{1}{(2\pi)^2}[\ta(-\de)\frac{d}{dU}(\ta(U-1)(\frac{U}{2}
\ln\frac{U+1}{U-1} - 1)) - \ta(-\de)\frac{d}{dV}(\ta(1-|V|)
\nonumber \\  &\times&(\frac{V+i\e}{2}\ln
\frac{V+i\e+1}{V+i\e-1} - 1)) -\frac{d}{dV}(\ta(V-1)(\frac{V}{2}
\ln\frac{V+1}{V-1} -1))]. 
\end{eqnarray}
This explicit expression indeed contains various combinations of the
logarithmical, inverse and delta-function's singularities at the 
light-cone and the reflected light-cone.

Let us take the boundary scaling limit $z' \rightarrow 0$ for the
 obtained Feynman propagator. The leading part of the 
bulk Feynman propagator for even $d$, when the spacelike separation is
chosen, is estimated as
\begin{equation}
iG_F(z,x;z',x')\simeq i\ta(\de+z^2)(-1)^\frac{d}{2}
\frac{\sin 2h_+\pi}{2\nu\sin\nu\pi}z'^{2h_+}G_{B\partial}(z,x;x')
\label{egf}\end{equation}
with $G_{B\partial}=\Gamma(2h_+)/(\Gamma(\nu)\pi^{d/2})
(z/(\de+z^2))^{2h_+}$, which is expressed as $iG_F\simeq i\ta(\de+z^2)
z'^{2h_+}G_{B\partial}/2\nu$. Thus by taking a large $V$ limit 
 we have extracted the bulk-boundary propagator 
$G_{B\partial}$ from the last term in (\ref{sum}), which arises from
$G^{(1)}$ alone. In this case the scaling limit $z' \rightarrow 0$ is
effectively the same as the large spacelike distance limit. 
When a large $U$ limit is taken the first term gives 
a scaling limit of $iG_F$ for the timelike separation
\begin{equation}
iG_F(z,x;z',x')\simeq i\ta(-\de-z^2)e^{i\nu\pi}
\frac{\sin 2h_+\pi}{2\nu\sin\nu\pi}z'^{2h_+}\tilde{G}_{B\partial}(z,x;x')
\label{bgf}\end{equation}
with $\tilde{G}_{B\partial}=\Gamma(2h_+)/(\Gamma(\nu)\pi^{d/2})
(z/(-\de-z^2))^{2h_+}$, where we have taken
account of an identity $\ta(-\de-z^2)\ta(-\de)=\ta(-\de-z^2)$.
It is also described by $iG_F\simeq i\ta(-\de-z^2)z'^{2h_+}
G_{B\partial}/2\nu$. This scaling limit is identical with
the large timelike interval limit. Whether the boundary scaling limit is
taken from the spacelike direction or the timelike direction, the Feynman
propagator shows the same behavior. The phase factor $e^{i\nu\pi}$ plays
a role to obtain the same expression. Furthermore, in the odd $d$ case
the last term in (\ref{gfo}) gives the expression (\ref{egf}) with 
$(-1)^{d/2}/\sin\nu\pi$ replaced by $(-1)^{(d-1)/2}/\cos\nu\pi$
  for the spacelike separation, from which we have again $iG_F\simeq 
i\ta(\de+z^2)z'^{2h_+}G_{B\partial}/2\nu$, while the first term provides
the expression (\ref{bgf}) with $i/\sin\nu\pi$ replaced by $-1/\cos\nu\pi$
for the timelike separation, which takes the form $iG_F\simeq 
i\ta(-\de-z^2)z'^{2h_+}G_{B\partial}/2\nu$. Though the expression of
the Feynman propagator depends on the dimension of AdS spacetime, we have
observed that the scaling behaviors are the same for the even and odd 
dimensions, as shown by 
\begin{equation}
iG_F \simeq i(\ta(\de+z^2) +\ta(-\de-z^2))z'^{2h_+}G_{B\partial}/2\nu
=iz'^{2h_+}G_{B\partial}/2\nu.
\label{fsc}\end{equation}
It is interesting that the two Heaviside functions happen to combine 
into unity. We shoud mention that this compact expression is presented
up to the delta-function terms. Thus from the explicit form of the
bulk Feynman propagator we derive the bulk-boundary propagator, however
an extra factor $1/2\nu$ is accompanied. The factor $1/2\nu =1/(4h_+ -d)$
was first emphasized in Ref. \cite{SG} by using an AdS variant of the 
Green theorem. It was argued in the classical 
evaluation of the action for a massive scalar
field in the Eulidean AdS space for the Poincar\'e coordinate \cite{KW}.
Recently in Ref. \cite{SBG} with respect to the global coordinate
the Feynman propagator in the Lorentzian $AdS_{d+1}$
spacetime has been investigated and from it the bulk-boundary propagator
has been extracted with the factor $1/2\nu$.

Here we analyse the short-distance property of the Hadamard function.
If the exact form of the symmetric function $G^{(1)}$ is known, it is 
possible to calculate the vacuum expectation value
of the square of scalar field. 
 For simplicity we will consider the $AdS_{2+1}$ spacetime. 
In order to extract the short-distance property we must expand $G^{(1)}$
around $V=1$. For the spacelike separation $\sigma >0$ with 
$V = 1 + \sigma$ the last two terms in (\ref{hae}) yield
$G^{(1)} \sim \ta(\sigma)/2\pi(1/\sqrt{2\sigma} - \nu
+ O(\sqrt{\sigma}) )$, where imaginary 
$Q^{\frac{1}{2}}_{\nu-\frac{1}{2}}$
itself is expanded. For the timelike separation 
$\sigma < 0$ the second term in (\ref{hae}) also gives $G^{(1)} 
\sim -\ta(-\sigma)\nu/2\pi$ where real 
$Q^{\frac{1}{2}}_{\nu-\frac{1}{2}}(V)$ for $|V|<1$ expressed in terms of
a hypergeometric function is expanded by using a relation between
the hypergeometric function with argument $V^2$ and that with $1-V^2$.
In this evaluation we have seen that the singular terms of the form
$1/\sqrt{2\sigma}$ are canceled out. But in analysing the Feynman 
propagator (\ref{sum}) with $d=2$, the additional  
$P^{\frac{1}{2}}_{\nu-\frac{1}{2}}$ term yields the singular term
$1/\sqrt{2\sigma}$ for the timelike separation. By subtraction of the
divergence in the coincidence limit for the Hadamard function 
$G^{(1)}\sim (\ta(\sigma)/\sqrt{2\sigma} - \nu)/2\pi$ we obtain
a regularized value $<0|\Phi^2|0>\sim -\nu/4\pi$,  whose negative
sign is also seen in the squared-mass term for the vacuum expectation
value of the stress tensor in $AdS_{1+1}$ for the global vacuum \cite{SS}

It was suggested that there is a simple relation $\mathcal{O}(x) =
\lim_{z\rightarrow 0}z^{-\Delta}\Phi(z,x)$ between a quantum scalar bulk
field $\Phi(z,x)$ and the corresponding dual boundary operator  
$\mathcal{O}(x)$ with conformal dimension $\Delta=2h_+$ \cite{BDHM,BGL}.
Here combining this relation with (\ref{fsc}) and taking the remaining $z$
to the boundary we have the conformal correlation function
\begin{equation}
<0|T(\mathcal{O}(x) \mathcal{O}(x')|0> \simeq \frac{1}{2\nu}
 \frac{\Gamma(\Delta)}{\Gamma(\nu)\pi^{d/2}(\de)^{\Delta}}.
\label{cfo}\end{equation}
For the AdS/CFT correspondence in the Euclidean $AdS_{d+1}$ space the
conformal two-point function is given by  $<\mathcal{O}_E(x)
\mathcal{O}_E(x')> = ((2\Delta-d)\Gamma(\Delta)/\pi^{d/2}
\Gamma(\Delta-d/2))|x-x'|^{-2\Delta}$, whose  extra factor $(2\Delta-d)$
was derived by putting the boundary at $z=\e$ and taking  
$\e$ to 0 \cite{GKP,MV}
and further proved to be consistent with the Ward identities \cite{FMMR}.
This factor has been argued from the different view point 
 in Ref. \cite{KW},
where the boundary behavior of the scalar field is expressed as 
$\Phi(z,x) \rightarrow z^{d-\Delta}\Phi_0(x) + z^{\Delta}A(x)$
where $\Phi_0(x)$ is a source function. The two-point function of the 
physical fluctuation $A(x)$, which rather corresponds to our conformal
field $\mathcal{O}(x)$, has been presented as   
$<A(x)A(x')>=(\Gamma(\Delta)/2\nu\pi^{d/2}\Gamma(\nu))|x-x'|^{-2\Delta}$,
where the normalization $A(x)=\mathcal{O}_E(x)/(2\Delta-d)$ is 
taken into account. This correlator in the Euclidean space for the 
Poincar\'e coordinate shows the same behavior as our obtained correlation
function (\ref{cfo}) in the Lorentzian spacetime.

By carrying out the mode integrations we have constructed the 
configuration forms of the Hadamard function and the Feynman propagator 
in the $(d+1)$-dimensional Lorentzian AdS spacetime for the Poincar\'e
coordinate. They are expressed in terms of an AdS-invariant distance
variable. The construction of the Hadamard function has been guided by
the experience with the direct evaluation of the retarded Green function.
Though the real retarded Green function is described by the (associated)
Legendre functions of the first and second kinds, we have observed
that the combined complex Feynman propagator is analytically described
by only those of the second kind. 

It has been shown that for the spacelike separation the 
retarded Green function is zero, while the Feynman propagator
indicates a power fall-off. This power behavior is compared with
the exponential tail of the Feynman propagator for the large
spacelike separation in the flat Minkowski spacetime. The power 
fall-off in the spacelike IR limit is intimately connected with the 
boundary scaling limit producing the bulk-boundary propagator.
In the timelike IR limit the Feynman propagator shows the same
power fall-off that is also associated with the boundary scaling
limit for the timelike separation.
In the higher-dimensional AdS spacetime, owing to the increase of 
the differential operations, the Feynman propagator becomes more
singular and more involved with a variety of singularities. 

We have seen that the evaluation of the Hadamard function depends
on the dimension of AdS spacetime, even or odd, corresponding
to that of the retarded Green function. For the bulk Feynman
propagator, which is characterized by the Heaviside functions
that divide the timelike and spacelike regions in the Lorentzian
AdS spacetime, we have taken one argument to the boundary and
exhibited that it reduces to the bulk-boundary propagator with a 
preferable factor $1/2\nu$. In this reduction the Heaviside function
dependence disappears and there is no difference between the 
even-dimensional case and the odd one. 
Moreover this factor is left intact in the 
two-point function of the dual boundary operator, when the other argument
is taken to the boundary.

Though we have estimated the vacuum expectation value of the 
square of scalar field, it would be interesting to study that
of the stress tensor from the obtained Hadamard function and
ask how differences appear between the even-dimensional AdS case
and the odd one. The position forms of the Feynman propagator
and the Hadamard function would provide important tools with which
to examine in greater detail the Lorentzian AdS/CFT correspondence
from the dynamical view point.

\end{document}